# Hub-Collision Avoidance and Leaf-Node Options Algorithm for Fractal Dimension and Renormalization of Complex Networks

Fei-Yan Guo,[1] Jia-Jun Zhou,[2] Zhong-Yuan Ruan,[2] Jian Zhang,[1] and Lin Qi[1,*]

[1)]School of Economics and Management, Beijing Information Science and Technology University, Beijing 100192, China

[2)]Institute of Cyberspace Security, Zhejiang University of Technology, Hangzhou 310023, China

(*Electronic mail: qilin@bistu.edu.cn.)

The box-covering method plays a fundamental role in the fractal property recognition and renormalization analysis of complex networks. This study proposes the hub-collision avoidance and leaf-node options (HALO) algorithm. In the box sampling process, a forward sampling rule (for avoiding hub collisions) and a reverse sampling rule (for preferentially selecting leaf nodes) are determined for bidirectional network traversal to reduce the randomness of sampling. In the box selection process, the larger necessary boxes are preferentially selected to join the solution by continuously removing small boxes. The CBB, MEMB, OBCA, and SM30 algorithms are compared with HALO in experiments. Results on 9 real networks show that HALO achieves the highest performance score, and obtains 11.40%, 7.67%, 2.18%, and 8.19% fewer boxes than the compared algorithms, respectively. The algorithm determinism is significantly improved. The fractal dimensions estimated by covering 4 standard networks are more accurate. Moreover, different from MEMB or OBCA, HALO is not affected by the tightness of the hubs and exhibits a stable performance in different networks. Finally, the time complexities of HALO and the compared algorithms are all $O(N^2)$, which is reasonable and acceptable.

**The fractal structure widely exists in various complex self-organized systems, such as the structure of protein molecules, the human social system, the urban**

**system, and its subsystems. Box-covering and renormalization analysis of fractal networks helps us to develop a deeper understanding of the structures and functions of complex systems. In this paper, the box-covering algorithm with bidirectional sampling rules is proposed. And we demonstrated the superior performance of the algorithm in terms of accuracy, efficiency, and stability.**

## I. INTRODUCTION

In recent decades, complex networks have been extensively studied to solve various practical problems in natural and social systems[1-5]. The exploration of network topology properties is an important aspect of complex network research[6-9]. Fractal and self-similar properties, which are the basic topological properties of complex networks[10], widely exist in various complex self-organizing systems[11-15] and have been applied in many fields[16-20]. Song et al. proposed a box-covering method to measure the fractal dimension of a network and realize renormalization, which is used to identify the fractal and self-similar properties of networks[10]. In this method, the box consists of a set of nodes in the network covered by it where the distances $l_{ij}$ between arbitrary two nodes $i$ and $j$ in the box is less than the box size $\ell_B$[10]. Given a box size $\ell_B$, the rule is to tile the entire network with the minimum number of boxes $N_B$. If the power-law relationship between $N_B$ and $\ell_B$, i.e.,

$$N_B \sim \ell_B^{-d_B} \tag{1}$$

has a definite fractal dimension $d_B$, the network has the fractal property[10, 21]. After the entire network is tiled by boxes, each box becomes a node, and all edges between the original nodes in the different boxes degenerate into edges between the new nodes, which is called the renormalization procedure of the network[22]. Hence, the box-covering method plays a fundamental role in the fractal property recognition and renormalization analysis of complex networks.

The ultimate goal of the box-covering algorithm is to find the optimal solution, i.e., to determine the minimum $N_B(\ell_B)$ value for any given box size $\ell_B$ [23]. However, since the box-covering problem belongs to the family of NP-hard problems[10, 24], existing algorithms choose to use optimization approximations to find an approximate solution for the optimal solution for an arbitrary value of $\ell_B$ [23]. The fewer the $N_B(\ell_B)$ value of the approximate solution, the closer it is to the optimal solution, which indicates that the approximate solution is more excellent. To determine the approximate solution in a limited time, scholars have proposed numerous box-covering algorithms following the greedy strategy[25], such as algorithms based on graph colouring[23, 26-28], algorithms based on breadth-first search[29-32], and algorithms based on meta-heuristic ideas[33-36]. However, there is a trade-off between accuracy and efficiency. The general methodology for box covering proposed by Wei et al. is of pioneering significance. This approach divides box covering into two processes: box sampling and box selection. It assumes that any nondeterministic box-covering algorithm can serve as a sampling strategy and determines the box selection priority based on two basic box overlapping patterns[37]. The methodology is well realized by the combination of the small-box-removal strategy with maximal box sampling (SM)[37]. Results show that the small-box-removal strategy considerably outperforms the classical algorithms such as the compact-box-burning (CBB) and the maximum-excluded-mass-burning (MEMB) algorithms. However, in this method, the randomness of box sampling is high, and the coverage effect of the solution greatly depends on the sampling density. As the sampling density increases, the coverage effect can be improved, but the computation time also increases significantly.

Therefore, scholars have been striving to develop a box sampling strategy with rules such that the sampling box contains the optimal coverage scheme as much as possible, thus reducing the randomness of the algorithm and locating a more excellent approximate solution within an acceptable time. Fractal networks have a self-similar hierarchy; hubs are usually in the inner hierarchy, and a large number of leaf nodes are in the external hierarchy[22, 38]. On this basis, some algorithms set hubs as seed nodes to produce as large boxes as possible, but at the same time, the leaf nodes at the edge will need a large number of small boxes to cover, or in other words, leaf nodes being contained in small boxes[23, 26, 28, 32]. Some algorithms follow the opposite idea, preferring to cover leaf nodes to prevent the edge box from becoming too small, but this may destroy the optimal coverage of hubs, and the solution might be far from the optimal one[38, 39]. It can be seen that when using the sampling rule to improve the efficiency of the algorithm, both the hubs and the leaf nodes have a considerable influence on solution accuracy. Therefore, during the box sampling process, both types of nodes must be considered for sampling to produce as many excellent boxes as possible and contain the optimal coverage scheme.

This study proposes the hub-collision avoidance and leaf-node options (HALO) algorithm. In the box sampling process, considering both the hubs and leaf nodes, a forward sampling rule (for avoiding hub collisions) and a reverse sampling rule (for preferentially selecting leaf nodes) are determined, and the network is traversed through the two rules to produce as many large boxes as possible. In the box selection process, the small-box-removal strategy is applied to preferentially select the larger necessary boxes for the solution. Through the above design, the trade-off between accuracy and efficiency in box covering is resolved, and the more excellent approximate solution is obtained within an acceptable time.

The rest of this paper is organized as follows: In Sec. II, the rationality and algorithm process of HALO are introduced. Section III explains the comparative experiments conducted on real networks and standard networks to verify the comprehensive advantages of HALO in terms of accuracy and efficiency. Finally, the research conclusions are given in Sec. IV.

## II. HALO ALGORITHM

Fractality in complex networks mainly originates from strong, effective repulsion between hubs, so the existence of hubs has a substantial impact on the box covering of a network[22, 40]. Selecting hubs as seed nodes for burning search ensures the generation of a large "compact box"[23], that is, a box that can contain the maximum possible number of nodes. The basic idea of burning search (breadth-first search) is to generate a box by growing it from one randomly selected node towards its neighborhood until the box includes the maximum possible number of nodes[23]. Furthermore, as hubs may be directly connected or share a large number of nodes, hub collisions, that is, selected hubs are distributed in the same box, should be avoided, otherwise, these large boxes generated by hubs will overlap greatly, and the solution will be far from the optimal one[23, 32]. In which, overlapping of boxes means that the two boxes cover at least one same node[38]. Considering the existence of hubs and avoiding hub collisions, MEMB preferentially selects the node $p$ with the maximum "excluded mass" as the central node to burn within the radius $r_B$ so that the distance $l_{ip}$ from any node $i$ in the box to the central node $p$ is not more than $r_B$[23]. The relationship between the box radius $r_B$ and the box size $\ell_B$ fulfills $r_B = (\ell_B - 1)/2$, and the excluded mass of a node is the number of uncovered nodes within a distance not more than $r_B$[23]. In terms of the number of boxes obtained, MEMB performs well and is presently deemed an ideal method[41]. However, considering the hubs alone may lead to falling into the "edge trap", where a large number of leaf nodes at the network edge require more small boxes to cover, thereby increasing the number of boxes needed to tile the entire network. As shown in Fig. 1(a), selecting seed nodes according to the excluded mass and allowing overlapping to generate compact boxes, which is critical to reducing the number of boxes[37, 38], two hubs (nodes 9 and 11) produce the two largest compact boxes (blue circles), but four small boxes (orange and green circles) are required to cover the leaf nodes at the edge. The resulting number of boxes needed to tile the entire network is 6, which is two more boxes than the optimal solution in Figs. 1(c) and 1(d).

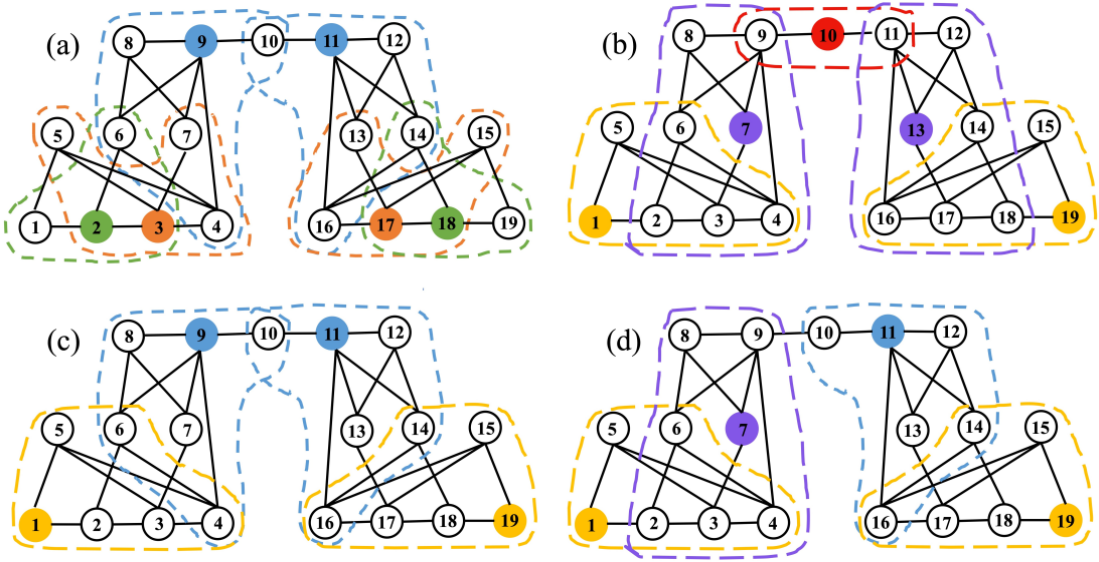

FIG. 1. (**a**) Illustration of the edge trap. At $r_B = 1$, according to the excluded mass, nodes 9 and 11 are preferentially selected as the central nodes to produce the largest compact boxes (blue circles); nodes 3 and 17 are selected as the central nodes to produce compact boxes (orange circles); nodes 2 and 18 are selected as the central nodes to produce the smallest compact boxes (green circles). The number of boxes needed to tile the entire network is 6. (**b**) Illustration of the centrifugal trap. At $\ell_B = 3$, according to the degree value, nodes 1, 10, and 19 are preferentially selected as the seed nodes to produce compact boxes (orange and red circles), and nodes 7 and 13 are selected as the seed nodes to produce compact boxes (purple circles). The number of boxes needed to tile the entire network is 5. (**c**) and (**d**) Optimal box-covering solutions. The large boxes in 1(a) and 1(b) are selected to be retained for optimal coverages. The number of boxes of optimal solution is 4.

As seen in Fig 1(a), in the box sampling process, the size of the edge boxes covering a large number of leaf nodes is also vital to the box covering of a network. Leaf nodes can be preferentially selected as seed nodes for box sampling to make the edge boxes as large as possible. Considering a large number of leaf nodes, the overlapping-box-covering (OBCA) algorithm preferentially selects the node with the minimum degree as the seed node to burn within the diameter $\ell_B$ so that the distance between any nodes in the box is less than $\ell_B$[38]. Eventually, there is a noticeable reduction in the number of boxes obtained by use of the OBCA in the order of nodes from small degrees to large degrees[38, 41]. However, due to the existence of hubs, burning search from leaf nodes may lead to falling into the "centrifugal trap", where the hubs are contained in a small box, and the low-degree nodes around hubs require additional boxes to cover, thereby increasing the number of boxes needed to tile the entire network. As shown in Fig. 1(b), selecting seed nodes according to the degree (from small to large) to tile the network, and allowing overlapping to produce compact boxes, leaf nodes 1 and 9 produce two excellent edge boxes (yellow circles), but the compact box (red circle) burning from node 10 contain hubs (nodes 9 and 11), and the nodes around these hubs require two additional boxes (purple circles) to cover, thereby increasing the number of boxes.

In summary, hubs and a large number of leaf nodes have an important impact on the box-covering results of a network. If the excellent large boxes in Figs. 1(a) and 1(b) are obtained by sampling and retained by selection, the optimal solution can be obtained, as shown in Figs. 1(c) and 1(d). Hence, to reduce randomness and time consumption of sampling, produce as many excellent boxes as possible, and avoid the edge and centrifugal traps considering the existence of hubs and a large number of leaf nodes, this study proposes the HALO algorithm, which is divided into two continuous processes: box sampling and box selection. First, the box sampling process consists of two stages: forward sampling and reverse sampling. In the forward sampling stage, a forward sampling rule (hubs-collision avoidance) is defined based on the excluded mass, and the rule is used to traverse the network to complete a sampling cycle. In the reverse sampling stage, a reverse sampling rule (leaf-node options) is defined based on degree values, and the rule is used to traverse the network to complete a sampling cycle. The sum set of the two sampling results is the final box sampling result. Then, in the box selection process, the overlapping boxes obtained by box sampling are selected based on the small-box-removal strategy[37], which preferentially selects the "necessary box"[31, 37], that is, a box that contains at least one node that does not belong to other boxes. The boxes retained in the box selection process constitute the box coverage solution of the HALO algorithm. The details of the box sampling and box selection processes are shown in Algorithms 1 and 2.

---

**Algorithm 1:** Process 1: Box sampling process of HALO

**Input:** graph $G = (V, E)$, $r_B$

**Output:** sampled boxes $B$

1      Initialize sampling boxes $B = \emptyset$;

     **/* Step 1: Forward rule box sampling */**

2      Initialize uncovered nodes $U = V$, seed nodes $C = \emptyset$;

3      **while** $U \neq \emptyset$ **do**

4          $box = \emptyset$;

5          Calculate excluded mass $em_i$ for the current non-seed nodes;

6          $p = argmax_{i \in \{V \setminus C\}} em_i$;      // get the node with maximal excluded mass

7          $C = C \cup \{p\}$;

8          $box = box \cup \{p\}$;

9          $box = box \cup \{i | \forall i \in U \wedge l_{ip} \leq r_B\}$;      // nodes with distance no more than $r_B$ from $p$

10        $U = U \setminus box$;

11        $B = B \cup \{box\}$;

12      **end**

     **/* Step 2: Reverse rule box sampling */**

13      Initialize uncovered nodes $U = V$, $\ell_B = 2r_B + 1$;

14      Calculate degree value $d_i$ for each node;

15      **while** $U \neq \emptyset$ **do**

16        $box = \emptyset$;

17        $p = argmin_{i \in U} d_i$;      // get the node with minimal degree value

18        $box = box \cup \{p\}$;

19        $R = \{i | \forall i \in U \wedge l_{ip} < l_B\}$;      // nodes with distance less than $l_B$ from $p$

20        **while** R $\neq \emptyset$ **do**

21            $N = U \cap R$;

22            **if** $N \neq \emptyset$ **then**

23               $n \leftarrow$ RandomSamping($N$);      // randomly sample a node from $N$

24            **else**

25               $n \leftarrow$ RandomSamping($R$);      // randomly sample a node from $R$

| 26 | **end** |
| 27 | $box = box \cup \{n\}$; |
| 28 | $R = R \setminus n$; |
| 29 | $K = \{i | \forall i \in U \wedge l_{in} < l_B\}$;      // nodes with distance less than $l_B$ from $n$ |
| 30 | $R = R \cap K$; |
| 31 | **end** |
| 32 | $U = U \setminus box$; |
| 33 | $B = B \cup \{box\}$; |
| 34 | **end** |
| 35 | **Return:** sampled boxes $B$ |

---

**Algorithm 2:** Process 2: Box selection process of HALO

**Input:** sampled boxes $B$

**Output:** box covering $Boxes$

| 1 | $S = \{|b_i| | \forall b_i \in B\}$;      // get box sizes |
| 2 | $B \leftarrow Sort(B, S)$;      // sort boxes in $B$ ascending in terms of box sizes |
| 3 | Calculate covering frequency $f[i]$ for each node $i$; |
| 4 | Initialize uncovered nodes $U = V$, $Boxes = \emptyset$; |
| 5 | **for** $j = 1$ to $|B|$ **do** |
| 6 |      **if** $\exists n \in b_j$ such that $f[n] = 1$ **then** |
| 7 |          $r = b_j \cap U$; |
| 8 |          $Boxes = Boxes \cup \{r\}$;      // cover the nodes in $r$ |
| 9 |          $U = U \setminus r$; |
| 10 |          $f[r] = 0$; |
| 11 |          $S = \{|b_i \cap U| | \forall b_i \in B\}$;      // update box sizes |
| 12 |          $B[j+1:] \leftarrow Sort(B[j+1:], S[j+1:])$; |
| 13 |      **else** |
| 14 |          $f[b_j] = f[b_j] - 1$; |
| 15 |      **end** |
| 16 | **end** |

As shown in Fig. 2, the sampling boxes in Fig. 2(a) contain all the excellent boxes shown in Figs. 1(a) and 1(b), and these excellent large boxes are retained through the box selection process of Fig. 2(b). The number of obtained boxes is the optimal solution, which is consistent with the results in Fig. 1(d). Therefore, HALO achieves the intended effect. The only difference between HALO and SM is the box sampling strategy. Compared with maximal box sampling of SM, bidirectional rules reduce the randomness and computation time of sampling and enable the generation of excellent large boxes. Then, through the box selection strategy, these large boxes are retained, thereby obtaining a more excellent approximate solution. Compared with MEMB and OBCA, HALO avoids both edge and centrifugal traps, which improves the effectiveness and applicability of the algorithm.

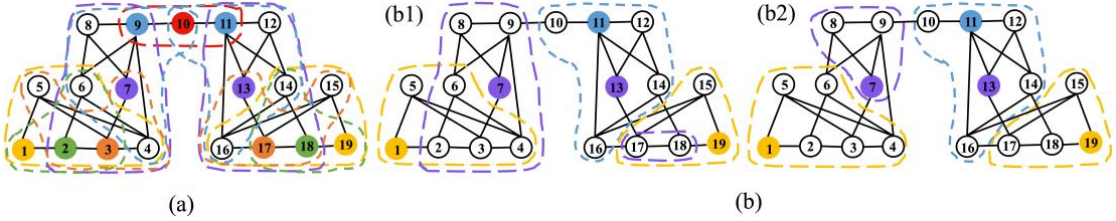

FIG. 2. Illustration of the HALO algorithm. (**a**) Box sampling. Forward traversal is performed within $r_B = 1$; nodes 9, 11, 3, 17, 2, and 18 are selected as the seed nodes to produce compact boxes (blue, orange, and green circles). Reverse traversal is conducted within $\ell_B = 3$; nodes 1, 10, 19, 7, and 13 are selected as the seed nodes to produce compact boxes (yellow, red, and purple circles). The number of sampling boxes is 11. (**b**) Box selection. The sampling box sizes are 3, 4, 5, 6, and 7. Small boxes of sizes 3, 4, and 5 (red, green, and orange circles in (**a**)) are removed, and the box with seed node 9 continues being removed until the box with the seed node 11 becomes a necessary box. The necessary box is selected, and the sizes of the remaining boxes are updated to obtain (b1). The sizes of the remaining boxes in (b1) are 2, 4, 6, and 7. Small box of size 2 is removed, boxes with seed nodes 1 and 19 are selected as necessary boxes and the remaining box is updated. The updated remaining box, with seed node 7, is also selected as the necessary box, and the box coverage is shown in (b2). Finally, the number of boxes needed to tile the entire network is 4.

## III. ALGORITHM PERFORMANCE EVALUATION

### A. Real network data

Comparative experiments were conducted on real networks to verify the effectiveness and applicability of HALO. Table I shows detailed information about the 9 analysed real networks, which are classified into four types, namely, biological, brain, social and economic networks; some of these networks have also been analyzed in related literature[10, 23, 30, 31, 38]. Some of the analyzed networks are directed, are weighted, contain more than one connected component or contain self-loops. In this study, the directed edges were converted to undirected edges, the weights were set to 1, the largest connected component was extracted and self-loops were removed to obtain an undirected, unweighted, connected network that suited the box-covering algorithm.

TABLE I. Real network data.

| Id | Network Code | Full Name | Description | References | Type |
|---|---|---|---|---|---|
| (a) | *S.cere* | Saccharomyces Cerevisiae Whole Network | Whole cellular network for Saccharomyces Cerevisiae | 42 | |
| (b) | *A.fulgidus* | Archaeoglobus fulgidus Whole Network | Whole cellular network for Archaeoglobus fulgidus | 43 | Biological network |
| (c) | *C.elegans* | Caenorhabditis Elegans Whole Network | Whole cellular network for Caenorhabditis Elegans | 44 | |
| (d) | *E.coli* | Escherichia Coli Whole Network | Whole cellular network for Escherichia Coli | 45 | |
| (e) | *mouse* | Mouse kasthuri graph v4 | Mouse brain network | 46 | Brain network |
| (f) | *twitch* | Musae twitch PTBR | Social networks of Twitch users | 47 | |
| (g) | *wiki* | Wiki Vote | Wikipedia who-votes-on-whom network | 46 | Social network |
| (h) | *facebook* | Facebook pages tvshow | Mutual likes among Facebook pages | 46 | |
| (i) | *poli* | Economic poli | Economic Network | 46 | Economic network |

The typical topological properties of the 9 real networks after processing are shown in Table II, where $N$ is the number of nodes, $E$ is the number of edges, $<k>$ is the average degree, $D$ is the network diameter, $M$ is the network modularity, $C$ is the clustering coefficient, $r$ is the assortativity coefficient, and $u$ is the rich-club coefficient. The value of $u$ is the ratio of the actual number of edges between the nodes constituting the top 5% of nodes (ranked by degree) to the maximum possible number of edges.

TABLE II. Typical topological properties of real networks.

| Id | $N$ | $E$ | $<k>$ | $D$ | $M$ | $C$ | $r$ | $u$ |
|---|---|---|---|---|---|---|---|---|
| (a) | 1812 | 4419 | 4.88 | 16.00 | 0.61 | 0.00 | -0.18 | 0.01 |
| (b) | 1557 | 3571 | 4.59 | 14.00 | 0.65 | 0.00 | -0.18 | 0.02 |
| (c) | 1461 | 3409 | 4.67 | 14.00 | 0.61 | 0.00 | -0.17 | 0.02 |
| (d) | 2859 | 6890 | 4.82 | 18.00 | 0.61 | 0.00 | -0.16 | 0.01 |
| (e) | 987 | 1536 | 3.11 | 12.00 | 0.63 | 0.00 | -0.24 | 0.00 |
| (f) | 1912 | 31299 | 32.74 | 7.00 | 0.30 | 0.34 | -0.23 | 0.40 |
| (g) | 889 | 2914 | 6.56 | 13.00 | 0.58 | 0.16 | -0.03 | 0.20 |
| (h) | 3892 | 17239 | 8.86 | 20.00 | 0.87 | 0.44 | 0.56 | 0.19 |
| (i) | 2343 | 2667 | 2.28 | 27.00 | 0.92 | 0.22 | -0.34 | 0.01 |

## B. Evaluation indicators and comparison algorithms

Four box-covering algorithms, namely, CBB, MEMB, OBCA, and SM, were compared with HALO in the experiments to verify the effectiveness and applicability of the proposed algorithm. In particular, the number of boxes $N_B$ for different box sizes $\ell_B$, performance score $S$, estimated fractal dimension $d_B$, and algorithm runtime $Time(s)$ were compared[25]. In the experiments described below, both the HALO algorithm and comparison algorithms ran 1000 times independently on 9 real networks, so the number of boxes $N_B$ and the running time $Time(s)$ at each box size $\ell_B$ are the means of 1000 independent experiments.

By examining the multi-index performance of existing box-covering algorithms in a large number of networks, scholars have divided these algorithms into three categories: algorithms with poor relative accuracy, algorithms with better accuracy but high time cost, and algorithms with ideal accuracy and efficiency[25]. CBB, MEMB, and OBCA belong to the third category, with OBCA being deemed the most accurate algorithm. SM belongs to the second category, but results show that it has the shortest runtime in this category. The runtime of SM depends on the sampling density $n$; a smaller $n$ reduces the runtime but affects accuracy. The original study on SM reports that a relatively stable, accurate result can be obtained when $n \geq 30$[39]. Therefore, in this study, HALO was compared with CBB, MEMB, OBCA, and SM to verify the comprehensive advantages of the proposed algorithm in terms of accuracy and efficiency. In the experiments, the sampling density $n$ of SM was set to 30, and the algorithm was named SM30.

## C. Result analysis

### 1. Comparative analysis of the number of boxes

First, the numbers of boxes $N_B$ obtained by each algorithm for different box sizes $\ell_B$ were compared. As shown in Fig. 3, HALO acquired fewer $N_B$ for different $\ell_B$. In particular, for small $\ell_B$, HALO obtained the minimum $N_B$ compared with the other algorithms, so it was closer to the optimal solution. For large $\ell_B$, HALO obtained the same $N_B$ as those of some of the compared algorithms, but it was still the minimum number of boxes in comparison. It follows that HALO can describe very well whether the network has the fractal property with smaller errors.

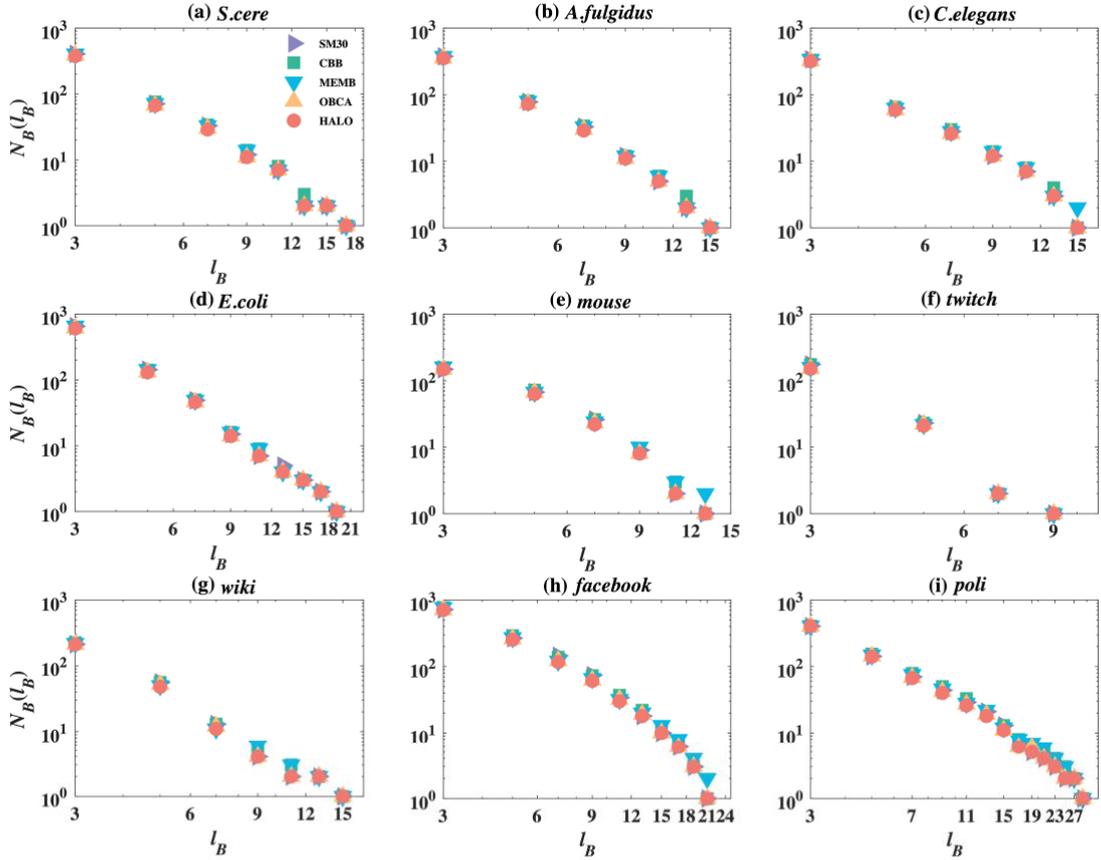

FIG. 3. The number of boxes $N_B$ comparison of HALO (red circle) with SM30 (purple right triangle), CBB (green square), MEMB (blue lower triangle), and OBCA (orange upper triangle). Shown are cases for (a) *S.cere*, (b) *A.fulgidus*, (c) *C.elegans*, (d) *E.coli*, (e) *mouse*, (f) *twitch*, (g) *Wiki*, (h) *Facebook*, and (i) *poli*.

Second, the average improvement ratio $avg(\eta)$ and highest improvement ratios $max(\eta)$ between $N_B$ produced by use of the HALO and the other algorithms were analyzed to further elucidate its advantage in terms of the number of boxes. The $avg(\eta)$ and $max(\eta)$ are calculated as follows：

$$avg(\eta) = avg((N_B^{CA} - N_B^{HALO})/N_B^{CA}) \qquad （2）$$

$$max(\eta) = max((N_B^{CA} - N_B^{HALO})/N_B^{CA}) \qquad （3）$$

where, $N_B^{CA}$ represents the mean of the results obtained by the comparison algorithms, and $N_B^{HALO}$ represents the mean of the results obtained by HALO[37, 38]. An appropriate box size range should be determined, and the $\eta$ values in this range must then be compared to obtain the suitable $avg(\eta)$ and $max(\eta)$. This is because, if $\ell_B$ is too large, $N_B$ will be significantly reduced. Even if the difference between $N_B^{HALO}$ and $N_B^{CA}$ is only 1, $\eta$ will be misleadingly large. For example, at a certain large $\ell_B$, $N_B^{CBB}$ is 3 and $N_B^{HALO}$ is 2. Then, $\eta_{CBB}$ can be as high as 33.33%. This situation was frequently observed in the experimental results of this study. Therefore, the box size where at least 10 boxes would be obtained was set as the lower limit of the box size range to eliminate the influence of tail noise and obtain the reasonable $avg(\eta)$ and $max(\eta)$[25]. As shown in Tables III and IV, the $avg(\eta)$ and $max(\eta)$ values between HALO and SM30, CBB and MEMB were considerable, with the mean values of $avg(\eta)$ being 8.19%, 11.40%, and 7.67%, and the mean values of $max(\eta)$ being 12.97%, 15.75%, and 11.51%, respectively. The values of $avg(\eta)$ and $max(\eta)$ between HALO and OBCA were relatively small (the mean values were 2.18% and 4.24%, respectively) and were affected by the box size. Mostly, for small $\ell_B$, HALO had better numbers of boxes compared with OBCA. However, due to the large size of the network, $N_B^{OBCA}$ and $N_B^{HALO}$ at small $\ell_B$ were large. Hence, although $N_B^{HALO}$ was significantly smaller than $N_B^{OBCA}$, $\eta_{OBCA}$ remained small, resulting in the relatively small $avg(\eta_{OBCA})$ and $max(\eta_{OBCA})$.

TABLE III. The average improvement ratio $avg(\eta)$ between the means of 1000 independent realizations of HALO and comparison algorithms.

| Id | $avg(\eta_{SM30})/\%$ | $avg(\eta_{CBB})/\%$ | $avg(\eta_{MEMB})/\%$ | $avg(\eta_{OBCA})/\%$ |
|---|---|---|---|---|
| (a) | 7.56 | 12.82 | 11.79 | 1.35 |
| (b) | 8.37 | 8.98 | 8.89 | 1.38 |
| (c) | 4.99 | 10.05 | 8.63 | 0.39 |
| (d) | 7.09 | 7.81 | 8.85 | 0.25 |
| (e) | 7.12 | 11.21 | 7.46 | 2.96 |
| (f) | 11.45 | 11.21 | 4.49 | 2.27 |
| (g) | 10.20 | 11.96 | 2.55 | 5.34 |
| (h) | 9.19 | 14.93 | 7.05 | 3.32 |
| (i) | 7.73 | 13.60 | 9.31 | 2.37 |
| mean value | 8.19 | 11.40 | 7.67 | 2.18 |

TABLE IV. The maximum improvement ratio $max(\eta)$ between the means of 1000 independent realizations of HALO and comparison algorithms.

| Id | $max(\eta_{SM30})/\%$ | $max(\eta_{CBB})/\%$ | $max(\eta_{MEMB})/\%$ | $max(\eta_{OBCA})/\%$ |
|---|---|---|---|---|
| (a) | 9.38 | 21.43 | 21.43 | 3.33 |
| (b) | 12.12 | 14.71 | 12.12 | 3.33 |
| (c) | 7.14 | 14.29 | 14.29 | 1.54 |
| (d) | 8.39 | 12.50 | 12.50 | 0.98 |
| (e) | 15.39 | 15.39 | 8.33 | 4.55 |
| (f) | 14.21 | 13.71 | 4.55 | 4.55 |
| (g) | 15.39 | 15.39 | 4.00 | 8.33 |
| (h) | 19.31 | 18.92 | 8.16 | 6.25 |
| (i) | 15.39 | 15.39 | 18.18 | 5.26 |
| mean value | 12.97 | 15.75 | 11.51 | 4.24 |

Third, Fig. 4 shows the probability distribution of $N_B$ at the same $\ell_B$. To reflect the multiscale improvement of HALO, the distribution results of networks in Fig. 4 corresponded to three box sizes in turn, namely, $\ell_B = 3$, $\ell_B = 5$, and $\ell_B = 7$. As shown in Fig. 4, the $P(N_B)$ of the HALO followed the Gaussian distribution. The distribution of $P(N_B)$ of the HALO had a smaller mean, and was steeper in most cases, that is, the algorithm results are more deterministic. The ratios $p$ of the coefficient of variation $V_S$ of HALO to those of the compared algorithms were calculated to determine the deterministic difference between HALO and the compared algorithms. The $p$ is calculated as follows:

$$p = V_s^{HALO}/V_s^{CA} \qquad (4)$$

where, $V_s^{CA}$ represents the coefficient of variation of the results obtained by the comparison algorithm, and $V_s^{HALO}$ represents the coefficient of variation of the results obtained by HALO. The $V_s$ is the ratio of the sample standard deviation to the mean, and it is generally used to compare the dispersions of different sample data. The smaller the $V_s$ is, the smaller is the data dispersion and the better the deterministic. As seen in Table V, the ratios of $p < 100.00\%$ were more common when $V_s = 0.00$ was disregarded, so HALO produced a smaller $V_s$. Compared with CBB and OBCA, HALO obtained a smaller $V_s$ at each $\ell_B$, and the minimum $p$ values were 21.73% and 70.48%, respectively. The $V_s$ of HALO was smaller than that SM30 in most cases, and the minimum $p$ was 43.07%. Particularly at small $\ell_B$, HALO outperformed MEMB, with a value of $p$ as small as 32.61%. Therefore, HALO obtains a smaller number of boxes and significantly increases the determinism of the results. In some of the networks in Fig. 4, the maximum values of HALO were better than the mean values of the compared algorithms and even better than the minimum value of the compared algorithms, which shows that even a single realization of HALO obtained reliable results. In Fig. 4, the performance of SM30 in many networks was obviously inferior to that of HALO. This is attributed to the high randomness of the box sampling of SM30; therefore, the result is not ideal despite the use of a reasonable box selection strategy. This shows that the sampling strategy for bidirectional rules in HALO is reasonable and effective.

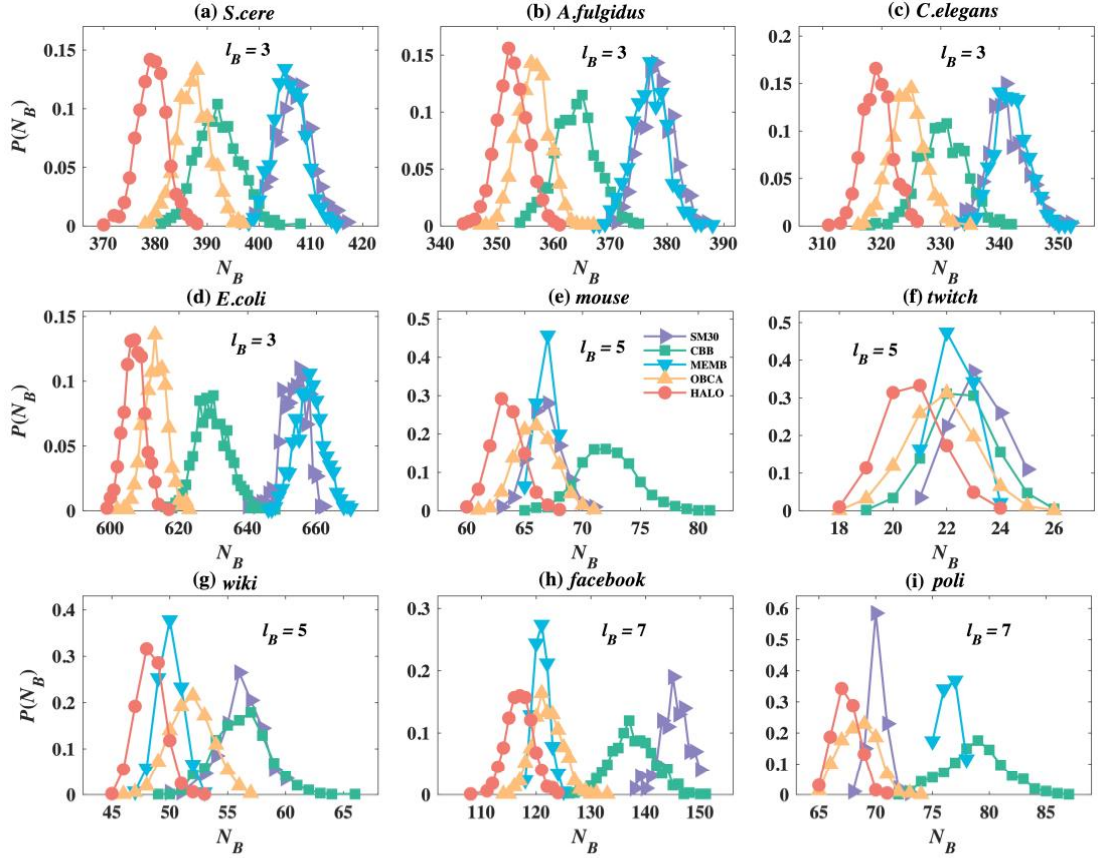

FIG. 4. The probability distribution $P(N_B)$ comparison of $N_B$ for 1000 independent realizations of HALO (red) with SM30 (purple), CBB (green), MEMB (blue) and OBCA (orange). Shown are cases for (a) *S.cere*, (b) *A.fulgidus*, (c) *C.elegans*, (d) *E.coli*, (e) *mouse*, (f) *twitch*, (g) *wiki*, (h) *facebook*, and (i) *poli*.

TABLE V. Ratio $p$ between the coefficients of variation of 1000 independent realizations of HALO and other algorithms. The results corresponded to three box sizes in turn, namely, $l_B = 3$, $\ell_B = 5$, and $\ell_B = 7$.

| | | Networks | | | | | | | | |
|---|---|---|---|---|---|---|---|---|---|---|
| | $p/\%$ | (a) | (b) | (c) | (d) | (e) | (f) | (g) | (h) | (i) |
| $\ell_B$ $= 3$ | $p_{\text{SM30}}$ | 87.23 | 98.43 | 78.75 | 83.98 | NA [a] | 43.07 | 437.38 | 96.20 | 0.00 |
| | $p_{CBB}$ | 70.00 | 74.60 | 66.61 | 62.39 | 0.00 [b] | 33.56 | 26.31 | 21.73 | 0.00 |
| | $p_{MEMB}$ | 101.01 | 95.61 | 93.17 | 76.27 | 0.00 | 87.70 | 51.69 | 32.61 | 0.00 |
| | $p_{OBCA}$ | 88.98 | 99.84 | 87.80 | 92.22 | NA | 90.22 | 93.53 | 90.02 | NA |
| $\ell_B$ $= 5$ | $p_{\text{SM30}}$ | 76.49 | 95.01 | 105.03 | 65.65 | 100.35 | 121.17 | 80.64 | 82.08 | NA |
| | $p_{CBB}$ | 46.91 | 43.18 | 43.58 | 50.17 | 64.80 | 99.30 | 60.24 | 52.32 | 31.64 |
| | $p_{MEMB}$ | NA | 84.42 | 82.40 | 87.77 | 173.03 | 160.75 | 118.34 | 120.97 | 86.51 |
| | $p_{OBCA}$ | 94.80 | 89.30 | 91.78 | 77.40 | 85.36 | 92.17 | 70.48 | 73.98 | 96.57 |
| $\ell_B$ $= 7$ | $p_{\text{SM30}}$ | 88.60 | 127.72 | 100.80 | 131.41 | 76.25 | NA | 69.22 | 112.20 | 156.10 |
| | $p_{CBB}$ | 69.08 | 73.88 | 64.10 | 63.89 | 67.18 | NA | 80.80 | 70.55 | 50.01 |
| | $p_{MEMB}$ | 144.99 | 174.91 | 195.69 | 156.70 | 153.79 | NA | 216.64 | 177.85 | 136.91 |
| | $p_{OBCA}$ | 93.13 | 90.82 | 99.44 | 87.97 | 84.82 | NA | 74.49 | 92.83 | 72.28 |

[a] NA indicates the case when the $V_s^{CA}$ is 0.00.
[b] 0.00 indicates the case when the $V_s^{HALO}$ is 0.00.

Together, Figs. 3, 4, Tables III, and IV show that compared with the other networks, in the social networks, the performance of OBCA at some small $\ell_B$ was poorer, whereas the performance of MEMB was improved. For example, as indicated in Fig. 4(g), the $P(N_B)$ of MEMB was better than that of OBCA. According to Tables III and IV, the $avg(\eta_{MEMB})$ and $max(\eta_{MEMB})$ values in the social networks were smaller compared with those in the other types of networks, but $avg(\eta_{OBCA})$ and $max(\eta_{OBCA})$ were larger. This should be related to the unique topological properties of social networks. A comparison of the topological properties of the networks in Table II suggests that the rich-club coefficient $u$ of social networks are significantly higher than those of other types of networks; the assortativity coefficient $r$ is also larger, particularly that of *facebook*. Hence, the tightness of the hubs can cause algorithms to fall into edge or centrifugal traps, thereby reducing the box-covering performance of MEMB or OBCA. The closer the hubs are connected, the easier it is to fall into the centrifugal trap, that is, the hubs become more likely to be contained in the same small box, thereby reducing the final box-covering effect; otherwise, it becomes easier to fall into the edge trap. By contrast, HALO achieved excellent results and was not affected by the topological properties of the networks. Hence, the sampling strategy for bidirectional rules can indeed avoid both edge and centrifugal traps so that HALO has a stable performance and is suitable for multiple types of networks.

In summary, in comparison with the other algorithms, HALO obtains fewer boxes, has better determinism, has more stable performance and is suitable for many types of networks.

*2. Comparative analysis of the performance score*

The previous section analyzed the improvements of the HALO algorithm over the 4 comparison algorithms in terms of the number of boxes, but the difference between the approximate solution obtained by the current algorithm and the optimal solution remains unknown. To identify how close the approximate solution obtained by the current algorithm is to the optimal solution, the performance score $S$ was calculated and compared. The definition is as follows:

$$S(\ell_B) = 1 - \frac{N_B(\ell_B) - N_B^{base}(\ell_B)}{N_B^{base}(\ell_B)} \tag{5}$$

where, $N_B(\ell_B)$ is the number of boxes obtained by the algorithm at $\ell_B$; $N_B^{base}(\ell_B)$ is the lower bound on the optimal solution at $\ell_B$. Since the optimal solution for any reasonably sized network cannot be determined, a lower bound on the optimal solution is used as a benchmark of the number of boxes for the gap comparison. The estimation idea of this lower bound is to construct a dual network $G'$ of the original network $G$, in which two nodes are connected if the distance between them in $G$ is greater than or equal to $\ell_B$[23]. By using the dual network, the box covering problem can be mapped to the vertex-coloring problem, that is, the minimum number of boxes required to tile the entire original network $G$ at the box size $\ell_B$ is equal to the minimum number of colors required to color the vertices in the dual network $G'$[23]. In addition, in the vertex coloring problem, the lower bound on the minimum number of colors is the size of the maximum clique of the network[48]. Therefore, the lower bound on the size of the maximum clique in the dual network $G'$, which is obtained by using the heuristic greedy algorithm[48], can be used as the lower bound on the optimal solution for the number of boxes in network $G$. According to Eq. (5), the closer the performance score $S$ is to 1, the closer the approximate solution obtained by the algorithm is to the optimal solution, indicating that the performance of the algorithm is better. When the performance score $S = 1$, the algorithm has obtained the optimal solution.

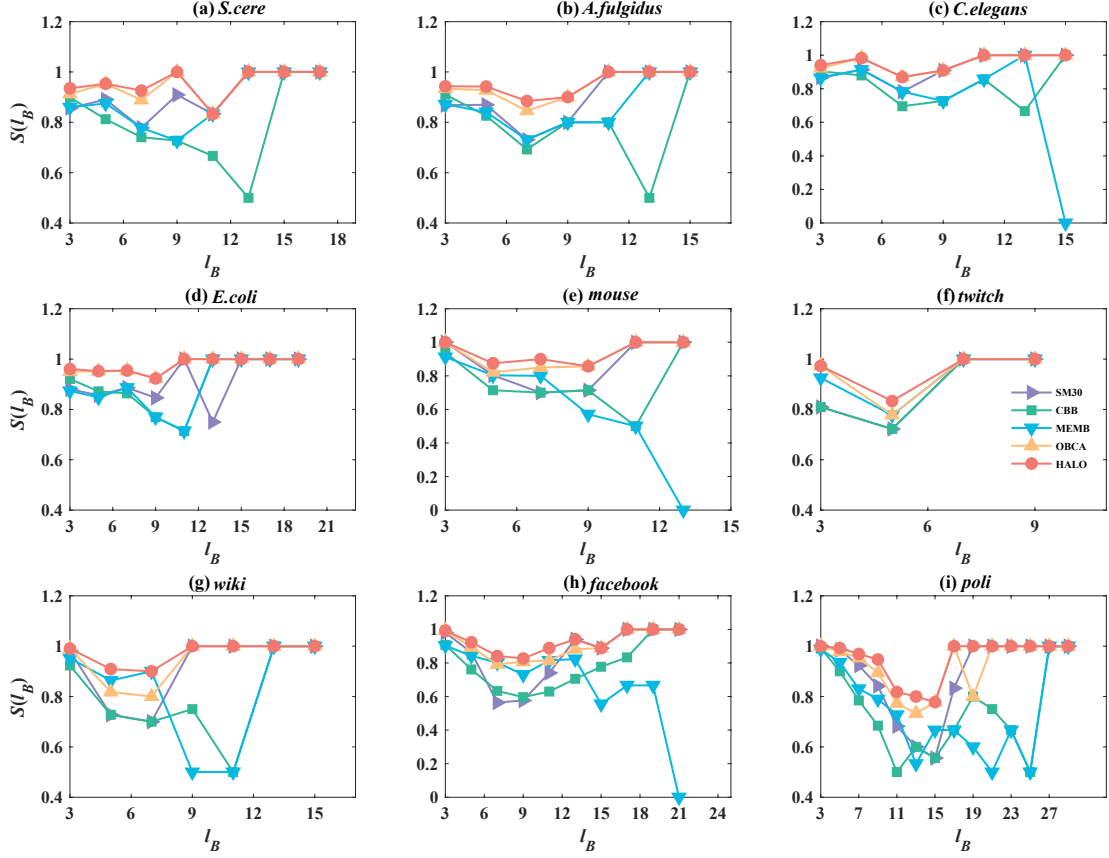

FIG. 5. The performance score $S$ comparison of HALO (red circle) with SM30 (purple right triangle), CBB (green square), MEMB (blue lower triangle), and OBCA (orange upper triangle). Shown are cases for (a) *S.cere*, (b) *A.fulgidus*, (c) *C.elegans*, (d) *E.coli*, (e) *mouse*, (f) *twitch*, (g) *wiki*, (h) *facebook*, and (i) *poli*.

Figure 5 shows the performance scores achieved by the HALO and 4 comparison algorithms, where $N_B(\ell_B)$ is the mean of 1000 independent realizations of the box-covering algorithm on the real network, and $N_B^{base}(\ell_B)$ is the maximum of 1000 independent realizations of the greedy algorithm on the dual network of the real network. It can be seen that HALO achieved the highest performance scores at different box sizes $\ell_B$, which is consistent with the results that HALO obtains the lowest number of boxes compared with comparison algorithms. In most cases, the performance scores of HALO were above 0.9, and at subsequent larger scales, the performance score reached 1, which indicates that HALO has achieved the optimal solution. It can be concluded that HALO delivers superior performance and can approximate or reach the optimal solution. In addition, when the other 4 algorithms have completed box coverage at all scales, the number of boxes of MEMB at the maximum box size in Figs. 5 (c), (e), and (h) were still 2, that is, the box coverage at all scales has not been completed, so the performance scores of MEMB at the maximum box size in these figures were 0.

*3. Estimation of fractal dimension*

The important application of the box-covering algorithm is to estimate the network fractal dimension. Therefore, the fractal dimensions obtained by HALO and 4 comparison algorithms on the network were compared. Eq. (1) can be converted to[30]:

$$\ln N_B(\ell_B) = -d_B \ln \ell_B + b \tag{6}$$

where, $d_B$ is the true fractal dimension and $b$ is a proper constant. According to Eq. (6), the number of boxes obtained by the algorithm and box size are linearly fitted in the double logarithmic coordinate to obtain the slope of the linear function which is the fractal dimension estimated by the algorithm.

TABLE VI. $d_B$ values estimated by HALO and the compared algorithms.

| Id | SM30 | CBB | MEMB | OBCA | HALO |
|----|------|------|------|------|------|
| (a) | 3.44 | 3.37 | 3.45 | 3.40 | 3.39 |
| (b) | 3.65 | 3.51 | 3.62 | 3.61 | 3.59 |
| (c) | 3.38 | 3.27 | 3.12 | 3.33 | 3.32 |
| (d) | 3.48 | 3.48 | 3.50 | 3.46 | 3.45 |
| (e) | 3.48 | 3.41 | 3.11 | 3.49 | 3.48 |
| (f) | 4.93 | 4.93 | 4.83 | 4.79 | 4.79 |
| (g) | 3.44 | 3.39 | 3.33 | 3.41 | 3.39 |
| (h) | 3.21 | 3.20 | 2.96 | 3.18 | 3.17 |
| (i) | 2.59 | 2.53 | 2.49 | 2.58 | 2.58 |

TABLE VI shows the fractal dimensions estimated by HALO and 4 comparison algorithms on 9 real networks, where the number of boxes $N_B$ used for fitting is the mean of 1000 independent realizations of the algorithm and the box size $\ell_B$ range is not restricted[25]. It can be seen that the estimated fractal dimension of a network varied with the applied box-covering algorithm, such as the fractal dimension of *mouse* changed from 3.11 (MEMB) to 3.49 (OBCA). Song et al. concluded that the fractal dimensions of *A.fulgidus*, *C.elegans*, and *E.coli* are all 3.5, and the fractal dimensions yielded by HALO were 3.59, 3.32, and 3.45, respectively[10]. The values between the two sets of results were close.

To further reflect the accuracy of HALO in estimating the fractal dimension, the standard network with known fractal dimensions[49, 50] can be selected for calculating and comparing the relative error of the fractal dimension. The fractal dimension relative error $\varepsilon$ is defined as:

$$\varepsilon = \left| \frac{d - d_B}{d_B} \right| \tag{7}$$

where, $d$ is the fractal dimension estimated by the algorithm, and $d_B$ is the true network fractal dimension[30]. The toroidal lattices with independently tuneable integer dimensions $D$ and degree values $K$ (without random rewiring) were chosen for experiments[49]. The parameters $K$ were selected as 6 degrees and 8 degrees, the parameters $D$ as 2 and 3 dimensions, and the number of nodes $N$ was determined as 5000. 4 standard networks were obtained by permuting and combining these three parameters. These networks were named *lattice-N-K-D* for box-covering experiments. Figure 6 shows the relative error $\varepsilon$ between the fractal dimensions estimated by HALO and 4 comparison algorithms running 100 times independently on 4 standard networks and the true fractal dimension, where the error line represents the standard deviation of multiple calculation results. Note that, since the influence of algorithm performance, MEMB obtained the same number of boxes at the larger box sizes in the tail when covering these toroidal lattices. Therefore, those tail noise points of MEMB were removed in the experiment to obtain the linear fitting results[25]. As can be seen from Fig. 6, the errors between the fractal dimension estimated by HALO and the true dimension were the smallest, both below 0.06, and the error fluctuation ranges were also small. These indicate that HALO can estimate the fractal dimension of the network more accurately, and can well describe whether the network has fractal characteristics.

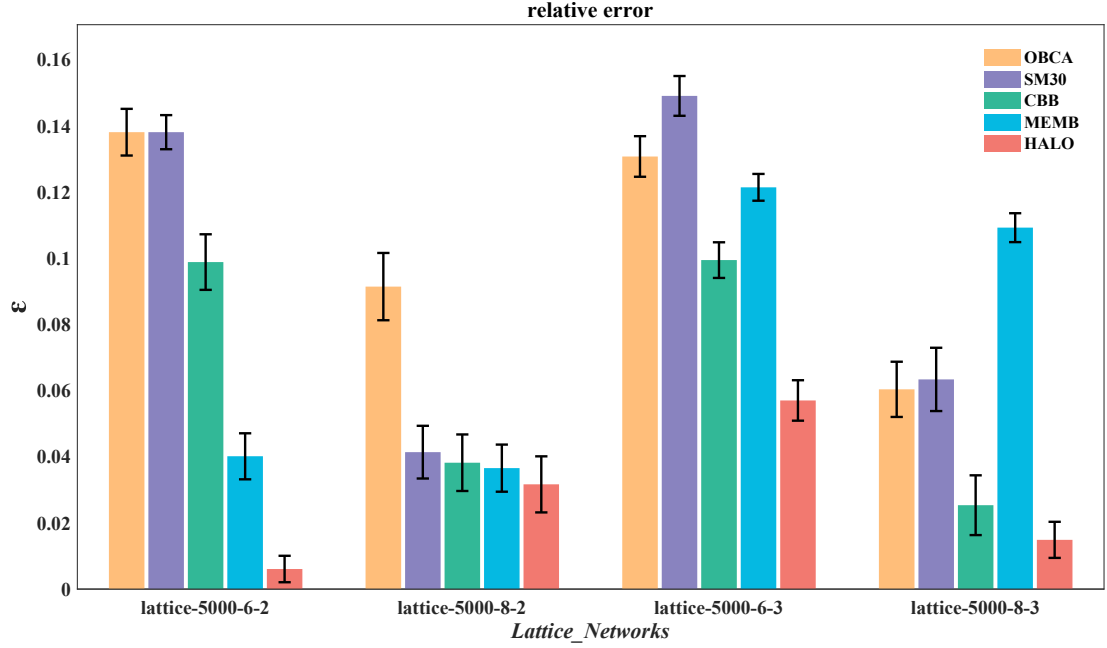

FIG. 6. The relative error $\varepsilon$ comparison of fractal dimension for 100 independent realizations of HALO (red) with SM30 (purple), CBB (green), MEMB (blue), and OBCA (orange). Shown are cases for *lattice-5000-6-2*, *lattice-5000-8-2*, *lattice-5000-6-3*, and *lattice-5000-8-3*.

## 4. Time performance analysis

The runtimes $Time\ (s)$ of HALO and 4 comparison algorithms at different box sizes were compared, and the relationship between $Time\ (s)$ and $\ell_B$ was plotted as shown in Fig. 7, where the y-axis is the logarithmic coordinate. With an increase in the size of $\ell_B$, the runtime of HALO tended to decrease. For most sizes $\ell_B$, the runtimes of HALO were much smaller than those of SM30, with the reduction ratio reaching 98.31%, but slightly higher than those of CBB, MEMB and OBCA. According to the implementation process, each algorithm was divided into two processes, namely, box sampling and box selection, to analyse the time complexities of the algorithms. Analysis suggests that the box sampling and box selection of CBB were performed synchronously and implemented through double-nested loops, so the time complexity of CBB was $O(N^2)$. HALO,

MEMB, OBCA and SM30 executed box sampling through double-nested loops with a time complexity of $O(N^2)$ and implemented box selection through simple loops with a time complexity of $O(N)$. Thus, the time complexities of HALO and the compared algorithms were all $O(N^2)$, and these algorithms belong to the same order. As also seen in Fig. 7, in most networks, the runtime of HALO was significantly different from that of CBB at $\ell_B \leq 9$, but it was close to that of MEMB or that of OBCA. At $\ell_B \geq 11$, HALO exhibited a similar time consumption to those of CBB, MEMB and OBCA. Therefore, in comparison with the other algorithms, HALO had a reasonable, acceptable runtime under the premise of obtaining fewer boxes.

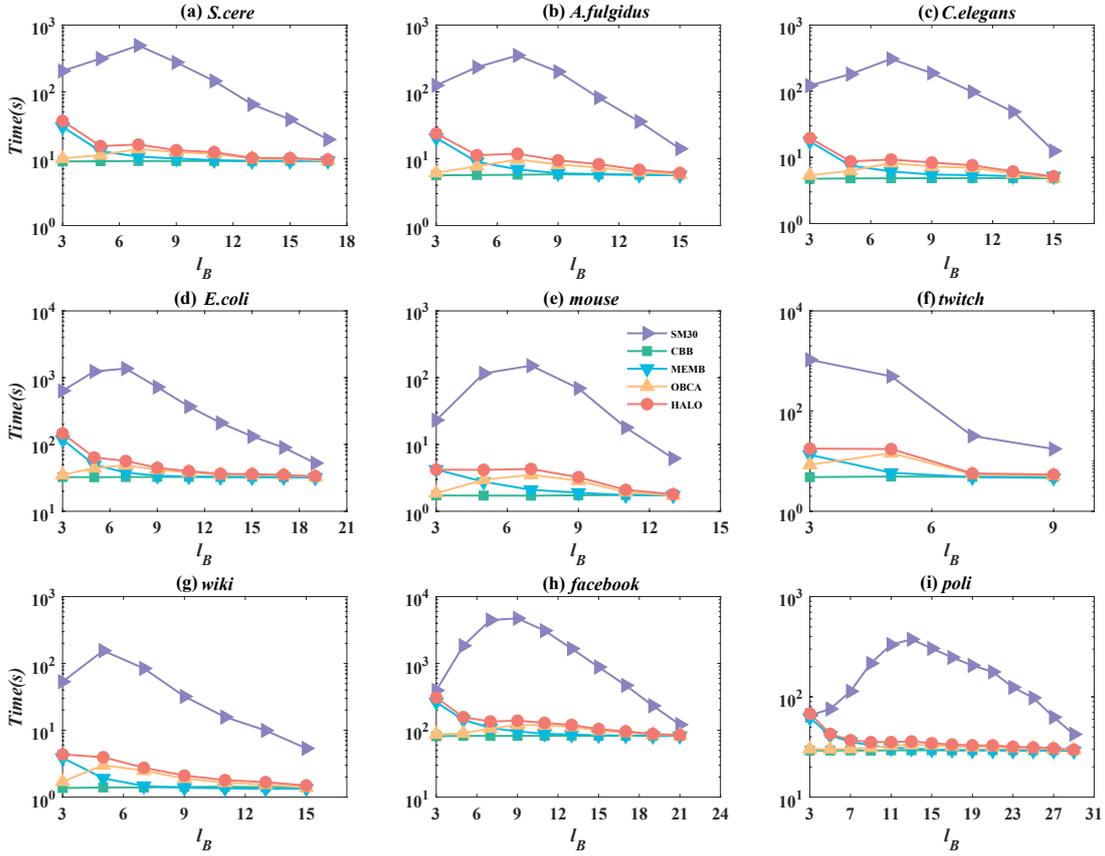

FIG. 7. Runtime $Time\ (s)$ comparison of HALO (red circle) with SM30 (purple right triangle), CBB (green square), MEMB (blue lower triangle), and OBCA (orange upper triangle). Shown are cases for (a) *S.cere*, (b) *A.fulgidus*, (c) *C.elegans*, (d) *E.coli*, (e) *mouse*, (f) *twitch*, (g) *wiki*, (h) *facebook*, and (i) *poli*.

# 4 CONCLUSIONS

To resolve the accuracy–efficiency trade-off encountered by current box-covering algorithms, this study proposes the HALO algorithm, which is divided into two processes: box sampling and box selection. In the box sampling process, a forward sampling rule (for avoiding hub collisions) and a reverse sampling rule (for preferentially selecting leaf nodes) are determined. This process reduces the randomness of sampling, which enables the sampling results to contain the optimal coverage as much as possible. In the box selection process, through the small-box-removal strategy, the large necessary boxes are preferentially selected to add them to the coverage scheme. Consequently, the algorithm obtained a more excellent approximate solution within an acceptable time. The advantages of the HALO over the CBB, MEMB, OBCA, and SM30 algorithms in terms of accuracy, stability, and efficiency are quantitatively analyzed.

In terms of accuracy, compared with CBB, MEMB, OBCA, and SM30, HALO obtains significantly fewer boxes, with the reduction ratios reaching 11.40%, 7.67%, 2.18%, and 8.19% on average, respectively. The determinism of the results is significantly improved compared with those of CBB, MEMB, OBCA, and SM30, with ratios of coefficients of variation reaching 21.73%, 32.61%, 70.48%, and 43.07%, respectively. The performance scores achieved by HALO are the highest, which indicates that the approximate solutions gained by HALO are closest to the optimal solutions. And on 4 standard networks, the fractal dimension relative errors achieved by HALO are the smallest, all below 0.06, which indicates that the HALO algorithm can estimate the network fractal dimension more accurately.

In terms of algorithm performance stability, HALO performs stably and well in various networks compared with MEMB and OBCA. The performance of MEMB and OBCA is affected by the tightness of hubs. Specifically, for some small $\ell_B$, OBCA is inferior to MEMB in social networks with tight hub connections and is unsuitable for social networks. The sampling strategy for bidirectional rules avoids both edge and centrifugal traps. Thus, the performance of HALO is not affected by the characteristics of network topologies, and its applicability is more extensive than that of MEMB and OBCA.

In terms of algorithm efficiency, the time complexities of HALO and the compared algorithms are all $O(N^2)$, which is reasonable and acceptable. Compared with the runtime of SM30, that of HALO is greatly reduced, with the maximum reduction ratio reaching 98.31%. At larger sizes (i.e., $\ell_B \geq 11$), the runtime is close to that of CBB, so HALO is suitable for large-scale networks.

In conclusion, HALO is superior to the current ideal algorithms. It can effectively estimate the fractal dimensions of networks and approach or reach optimal solutions of the box covering in an acceptable time. In addition, the sampling strategy for bidirectional rules fulfills its intended role, and is reasonable, effective, and necessary. The advantages of HALO in the three above-mentioned aspects give it application significance in real scenarios, especially analyses of real networks with unpredictable structures.

## DATA AND MATERIALS AVAILABILITY

The data and code that support the findings of this study are openly available on GitHub at https://github.com/guo2877/box-covering-algorithm.

## ACKNOWLEDGEMENTS

This study is supported by the National Key Research & Development Plan Project (2021YFF0901303), National Natural Science Foundation of China (72002017), National Social Science Fund of China (20BGL094), Youth Talent Promotion Program of Beijing Association for